# Synthesis and Superconducting Properties of a Hexagonal Phosphide ScRhP

Takumi Inohara[1], Yoshihiko Okamoto[1,3,*], Youichi Yamakawa[2,3], and Koshi Takenaka[1]

[1]*Department of Applied Physics, Nagoya University, Nagoya 464-8603, Japan*
[2]*Department of Physics, Nagoya University, Nagoya 464-8602, Japan*
[3]*Institute for Advanced Research, Nagoya University, Nagoya 464-8601, Japan*

We report the synthesis and superconducting properties of the ternary phosphide ScRhP. The crystal structure of ScRhP is determined to be the ordered $Fe_2P$ type with the hexagonal $P\bar{6}2m$ space group by powder X-ray diffraction experiments. Resistivity, magnetization, and heat capacity data indicate that ScRhP is a bulk superconductor with a transition temperature $T_c$ of 2 K. This $T_c$ is lower than that of its $5d$ analogue, ScIrP ($T_c$ = 3.4 K), although ScRhP is found to have larger electronic density of states at the Fermi energy and a higher Debye temperature than those of ScIrP.

## 1. Introduction

Many ternary pnictides with the general formula $MM'X$ ($X$ = P, As) crystallize in the ordered $Fe_2P$-type structure with the hexagonal $P\bar{6}2m$ space group, as shown in Fig. 1(a).[1] In general, $M$ is Ca, Sc, or early transition metal elements. $M$ atoms mainly occupy the 3g positions, square-pyramidally coordinated by $X$ atoms and forming a kagome-triangular lattice.[2] On the other hand, $M'$ is usually late transition metal elements with a smaller atomic radius than that of the $M$ atom. $M'$ atoms mainly occupy the 3f positions, tetrahedrally coordinated by $X$ atoms, and form $M'_3$ clusters with a regular-triangular shape, which form a triangular lattice. In many cases, $M$ and $M'$ atoms completely occupy 3g and 3f positions, respectively. However, the disorder of $M$ and $M'$ atoms may occur when they have similar atomic radii and electronegativities, as seen in NiCoP or FeRhAs.[3,4]

Superconductivity in $MM'X$ has long been a subject of study. The most intensively studied compounds are ZrRuP and its analogues, ZrRuAs and HfRuP. They show a superconducting transition at a relatively high $T_c$ of ~ 12 K.[5–10] ZrRuP has an orthorhombic polymorph with the ordered $Co_2P$-type structure, which is superconducting below 4 K, significantly lower than that of the hexagonal structure.[11,12] A theoretical study suggested that the charge-density-wave instability plays an important role in the occurrence of a relatively high $T_c$ in the hexagonal polymorph.[13] This instability does not exist in the orthorhombic polymorph. Moreover, MoNiP is reported to show superconductivity with an onset transition temperature of 15.5 K.[14,15]

Ordered $Fe_2P$-type pnictides are also promising candidates for realizing unconventional superconductivity. Since the ordered $Fe_2P$-type structure does not have space inversion symmetry, spin triplet superconductivity can sig-nificantly hybridize with spin singlet superconductivity when the compound contains a heavy element with strong spin–orbit interactions such as those in $CePt_3Si$[16] and $Li_2Pt_3B$.[17] A recently discovered superconductor, ScIrP, has been shown to have a relatively high upper critical field of $\mu_0 H_{c2}(0) = 5.11$ T, compared to its transition temperature of $T_c = 3.4$ K.[18] This $H_{c2}(0)$ may be enhanced by the strong spin–orbit interaction of the Ir 5$d$ electrons. In addition, CaAgP and CaAgAs are found to be a line-node Dirac semimetal with a Fermi ring centered at the Γ point in $k$-space and a topological insulator without inversion symmetry, respectively.[19] These results indicate that materials with the ordered $Fe_2P$-type structure are intriguing in the physics of their topological phases.

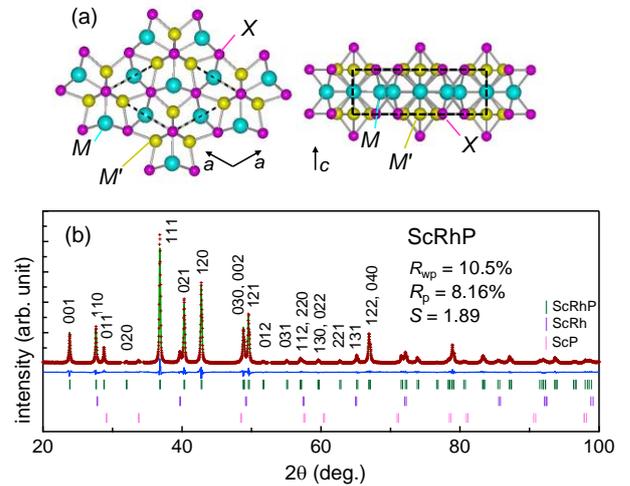

Fig. 1. (a) Crystal structure of $MM'X$ ($X$ = P, As) crystallizing in the ordered $Fe_2P$-type structure, viewing along (left) and perpendicular (right) to the $c$-axis. Large, medium, and small spheres represent $M$, $M'$, and $X$ atoms, respectively. In ScRhP, Sc, Rh, and P atoms occupy $M$, $M'$, and $X$ sites, respectively. The dashed line represents the hexagonal unit cell. (b) Powder XRD pattern of a ScRhP polycrystalline sample taken at room temperature. The crosses are the experimental data. Peak indices at $20° < 2θ < 70°$ are given using a hexagonal unit cell with lattice constants of $a$ = 6.4530(11) Å and $c$ = 3.72716(6) Å. The curve on the experimental data shows a calculated pattern and the lower curve shows a difference plot between the experimental and calculated intensities. The upper, middle, and lower vertical bars indicate the positions of the Bragg reflections of the ScRhP, ScRh, and ScP phases, respectively. Reliability factors in the Rietveld refinement are also shown.

In this paper, we report the synthesis and superconducting and normal-state properties of ScRhP, which is a new member of the ordered $Fe_2P$-type pnictide family. There is no previous study on the synthesis of ScRhP. We succeeded in preparing polycrystalline samples of ScRhP by a solid-state reaction method and found that ScRhP crystallizes in the ordered $Fe_2P$-type structure by employing the Rietveld analysis of powder X-ray diffraction (XRD) data. These samples show a bulk superconducting transition at $T_c = 2$ K. This $T_c$ is lower than those of its isoelectronic and isostructural compounds, ScIrP ($T_c = 3.4$ K) and ZrRuP ($T_c \sim 12$ K). We discuss the characteristic features of the superconductivity in ScRhP via comparison with ScIrP.

## 2. Experimental Procedure

Polycrystalline samples of ScRhP were prepared by a solid-state reaction method. Considering the low reactivity of Rh, a 0.9 : 1 : 1 molar ratio of Sc chip, Rh powder, and black phosphorus powder was mixed and sealed in a quartz tube with ~0.05 MPa of Ar gas. The tube was slowly heated to and kept at 673 K for 12 h and then at 1173 K for 48 h. The obtained sample was pulverized and sealed in a quartz tube with ~0.05 MPa of Ar gas. The tube was heated to and kept at 1173 K for 48 h. The obtained sample was pulverized again, pressed into pellets, wrapped in a Ta foil, and then sealed in a quartz tube with ~0.05 MPa of Ar gas. The tube was heated to and kept at 1423 K for 72 h.

The crystal structure of ScRhP was determined by the Rietveld analysis of powder XRD data, which was taken at room temperature using a RINT-2100 diffractometer with Cu Kα radiation (Rigaku), using the Rietan-FP program.[20] Magnetization measurements between 1.8 and 300 K were recorded using the magnetic properties measurement system (Quantum Design). Electrical resistivity and heat capacity measurements down to 0.5 K were recorded with the physical properties measurement system (Quantum Design). The first principles calculations for ScRhP and ScIrP were performed including spin–orbit coupling. The experimental structural parameters were used for the calculations. We used the full-potential linearized augmented plane-wave method within the generalized gradient approximation as implemented in the WIEN2k code.[21] In the self-consistent calculation, the $k$-point sampling of the Brillouin zone was set to 24 × 24 × 36 and the energy convergence criterion was 0.01 mRy.

Table I. Crystallographic parameters for ScRhP determined by powder XRD analysis. The space group is $P$–$62m$. The lattice constants are $a = 6.4530(11)$ Å and $c = 3.72716(6)$ Å. $g$ and $B$ are the occupancy and thermal displacement parameter, respectively. The $B$ values of the P1 and P2 sites are constrained to be the same.

|    |    | $x$       | $y$ | $z$ | $g$ | $B$ (Å$^2$) |
|----|----|-----------|-----|-----|-----|-------------|
| Sc | 3g | 0.5917(3) | 0   | 1/2 | 1   | 0.74(4)     |
| Rh | 3f | 0.2498(11)| 0   | 0   | 1   | 0.62(11)    |
| P1 | 1b | 0         | 0   | 1/2 | 1   | 0.57(5)     |
| P2 | 2c | 1/3       | 2/3 | 0   | 1   | 0.57(5)     |

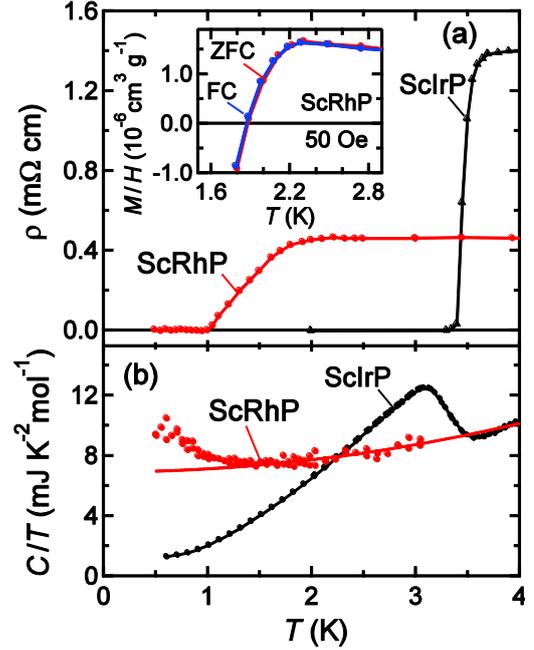

Fig. 2. (a) Temperature dependences of the electrical resistivities of polycrystalline samples of ScRhP and ScIrP. The inset shows the field-cooled and zero-field-cooled magnetizations of a ScRhP polycrystalline sample measured at a magnetic field of 50 Oe. (b) Temperature dependences of heat capacity divided by temperature for polycrystalline samples of ScRhP and ScIrP. The solid curve on the ScRhP data shows a fit to the equation $C/T = \beta T^2 + \gamma$ in the 2–5 K range.

## 3. Results and Discussion

### 3.1 Crystal structure

Figure 1(b) shows the powder XRD pattern of a ScRhP polycrystalline sample taken at room temperature. All diffraction peaks, except some small peaks of tiny amounts of ScRh, ScP, and unknown impurities, can be indexed on the basis of a hexagonal unit cell with the lattice constants of $a \sim 6.45$ Å and $c \sim 3.73$ Å, as shown in Fig. 1(b). This result indicates that the ScRhP phase is obtained as a main phase and crystallizes in the $Fe_2P$-type structure.

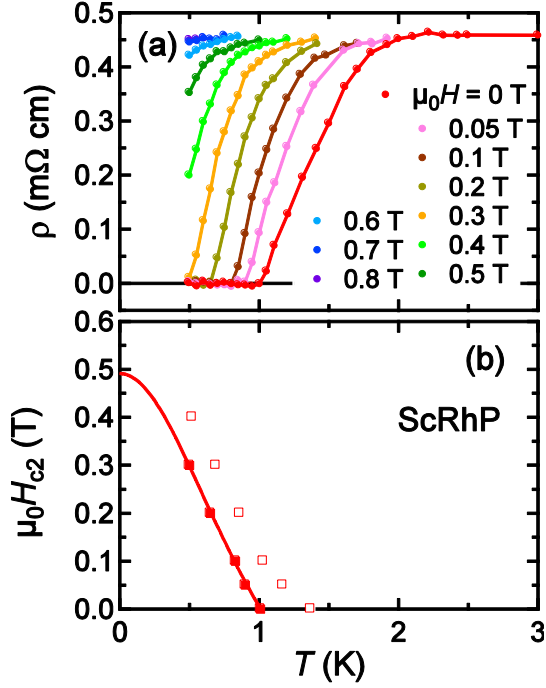

Fig. 3. (a) Temperature dependences of the electrical resistivity of a ScRhP polycrystalline sample measured at various magnetic fields of 0–0.8 T. (b) Temperature dependence of the upper critical field $H_{c2}$. Open and filled squares represent $H_{c2}$ values determined by the midpoint and zero-resistivity $T_c$, respectively. The solid curve on the latter shows a fit to the Ginzburg–Landau formula.

We performed the Rietveld analysis of the powder XRD data to determine whether the hexagonal $Fe_2P$-type structure is valid and whether the site disorder occurs in the ScRhP sample. The results of the Rietveld refinement using the $Fe_2P$-type structural model with the $P-62m$ space group are shown in Fig. 1(b) and Table I. The refinement of the structure by the model with other space groups with hexagonal symmetry gave worse results than that by the model with the $P-62m$ space group, suggesting that the $Fe_2P$-type structural model with the $P-62m$ space group is more valid. Moreover, regardless of the initial conditions considering the intersite defects or vacancies of Sc and Rh atoms, i.e., even if we put Sc and Rh atoms at 3f and 3g sites, respectively, or put them randomly at both sites, the refinements converged in a model in which the 3g and 3f sites are fully occupied by Sc and Rh atoms, respectively. This result suggests that ScRhP crystallizes in the ordered $Fe_2P$-type structure without site mixing occurring such as that in NiCoP and FeRhAs.[3,4] Hence, the structural parameters refined with the occupancies fixed to be one are shown in Table I. The lattice parameters obtained are $a = 6.4530(11)$ Å and $c = 3.72716(6)$ Å, which are 1–2% longer and ~4% shorter than the reported $a$ and $c$ of ScIrP, respectively.[18,22] As a result, the ratio of $c$ to $a$ of 0.580 in ScRhP is ~5% smaller than that in ScIrP ($c/a = 0.61$), indicating that the crystal structure of ScRhP is highly compressed along the $c$-axis, compared with that of ScIrP.

*3.2 Superconducting properties*

Figure 2 shows the temperature dependences of electrical resistivity, magnetization, and heat capacity divided by the temperature of a ScRhP polycrystalline sample. The electrical resistivity $\rho$ of the ScIrP polycrystalline sample is also shown for reference. The $\rho$ of the ScRhP sample exhibits a sudden drop to zero between 2.1 and 1.0 K with decreasing temperature. The magnetization $M$ measured at a magnetic field of $H = 50$ Oe slightly increases with decreasing temperature, suddenly decreases below 2.2 K, and then becomes negative at the lowest measured temperature of 1.8 K, as shown in the inset. Thus, the anomalies at around 2 K observed in the $\rho$ and $M$ data of the ScRhP sample are due to a superconducting transition and the negative $M$ at 1.8 K is a Meissner signal, although it is quite small because the lowest measured temperature is just below $T_c$.

The heat capacity divided by the temperature $C/T$ of the ScRhP sample, shown in Fig. 2(b), deviates upward below 1.5 K from the solid curve, which is the summation of the lattice contribution described by the Debye law and the electron contribution with a constant value, expected to appear in a normal metal at low temperatures. Then, $C/T$ seems to take a maximum value at around 0.6 K. This behavior corresponds to the superconducting transition observed in the $\rho$ and $M$ data, indicating that a bulk superconducting transition occurs in the ScRhP sample. Note that the resistivity drop at the superconducting transition of the ScRhP sample is much broader than that of ScIrP, as seen in Fig. 2(a). This suggests that the $T_c$ in the ScRhP sample is not uniform, possibly owing to the surface effects. Considering that the onset and midpoint of the resistivity drop are at 2.1 and 1.4 K, respectively, the onset of the magnetization drop is at 2.2 K, and the onset of the heat capacity jump is at 1.5 K, the $T_c$ of ScRhP is determined to be 2 K. The bulk nature of the superconducting transition should be confirmed more solidly by future experiments such as magnetization measurements using a $^3$He refrigerator and heat capacity measurements using a dilution refrigerator.

The electrical resistivity of the ScRhP polycrystalline sample measured at various magnetic fields is shown in Fig. 3(a). The resistivity drop due to the superconducting transition is suppressed below the lowest measured temperature of 0.5 K by applying a magnetic field of 0.8 T. As described below, this field dependence yields a short coherence length of ~20 nm, suggesting that ScRhP is a type-II superconductor. The upper critical fields $H_{c2}$ of ScRhP, determined by the midpoint and zero-resistivity $T_c$ in the $\rho$ data, are shown as a function of temperature in Fig. 3(b).

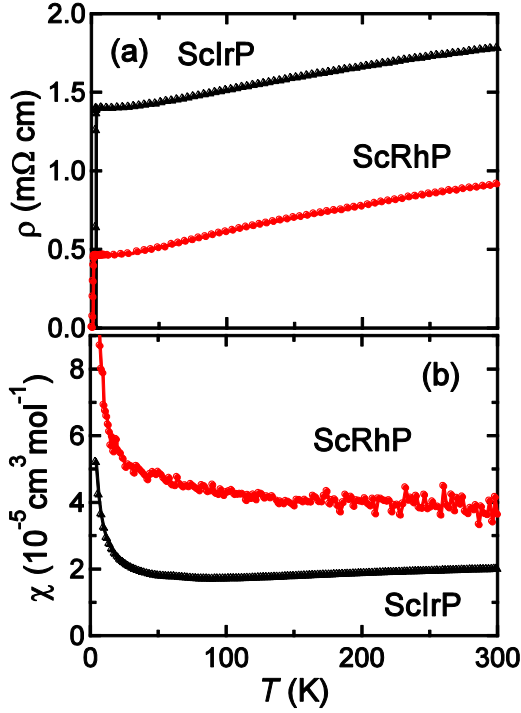

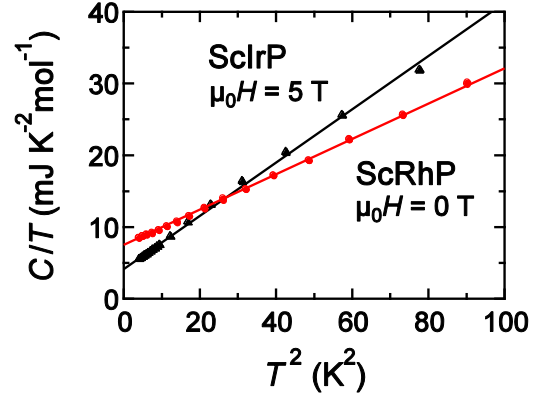

Fig. 5. Heat capacity divided by the temperature of ScRhP and ScIrP polycrystalline samples as a function of $T^2$. The data measured at magnetic fields of 0 and 5 T are shown for ScRhP and ScIrP, respectively. The solid lines show the results of linear fits of the low-temperature data for each sample.

Fig. 4. Temperature dependences of (a) electrical resistivity and (b) magnetic susceptibility measured at a magnetic field of 1 T for polycrystalline samples of ScRhP and ScIrP.

The midpoint $H_{c2}$ exhibits a concave-upward temperature dependence in the low-magnetic-field region below 0.1 T. This dependence reflects the broad superconducting transitions at low magnetic fields, as seen in Fig. 3(a).

As shown in Fig. 3(b), the $H_{c2}$ values determined by the zero-resistivity $T_c$, which is more intrinsic for the present sample than those determined by the midpoint $T_c$, are fitted to the Ginzburg–Landau (GL) formula $H_{c2}(T) = H_{c2}(0)[1 - (T/T_c)^2]/[1 + (T/T_c)^2]$. This fit yields $\mu_0 H_{c2}(0) = 0.491(6)$ T and a GL coherence length of $\xi_{GL} = 25.9$ nm. The $H_{c2}(0)$ estimated using the midpoint $T_c$ must be higher than 0.491 T, but should not exceed 0.6 T. Hence, $H_{c2}(0)$ is determined to be $\mu_0 H_{c2}(0) = 0.5$ T. This $H_{c2}(0)$ is higher than those of some Rh phosphide superconductors with similar $T_c$ values; $\mu_0 H_{c2}(0) = 0.5$ T of ScRhP is higher than that of not only centrosymmetric BaRh$_2$P$_2$ [$T_c = 1.0$ K, $\mu_0 H_{c2}(0) = 0.1$ T][23] but also noncentrosymmetric LaRhP [$T_c = 2.5$ K, $\mu_0 H_{c2}(0) = 0.27$ T].[24] However, the $H_{c2}(0)$ of ScRhP is only ~1/10 of that of the 5$d$ analogue ScIrP, although the $T_c$ of ScRhP is almost half that of ScIrP. The lower $H_{c2}(0)$ of ScRhP than that of ScIrP may correspond to the weaker spin–orbit interaction of the Rh 4$d$ electrons compared to the Ir 5$d$ electrons. This also implies that the $H_{c2}(0)$ in ScIrP is significantly enhanced by the strong antisymmetric spin–orbit interaction of the Ir 5$d$ electrons in the noncentrosymmetric crystal structure.

### 3.3 Electronic density of states

The physical properties of ScRhP in the normal state are discussed in comparison to those of ScIrP. Figure 4(a) shows the temperature dependences of the $\rho$ values of polycrystalline samples of ScRhP and ScIrP. Both samples show metallic behavior, where $\rho$ decreases with decreasing temperature, although their residual resistivities are significantly different. ScRhP and ScIrP have a finite Sommerfeld coefficient $\gamma$, consistent with the metallic $\rho$. Figure 5 shows the $C/T$ of ScRhP and ScIrP polycrystalline samples as a function of $T^2$. For ScIrP, the $C/T$ data taken at $\mu_0 H = 5$ T, in which the superconducting transition is suppressed below 2 K, are shown. A linear fit of the ScRhP data between 2 and 7 K to the equation $C/T = \beta T^2 + \gamma$, where $\beta$ represents the coefficient of the $T^3$ term of the lattice heat capacity, yields 0.246(15) mJ K$^{-4}$ mol$^{-1}$ and $\gamma = 7.52(4)$ mJ K$^{-2}$ mol$^{-1}$. The $\beta$ and $\gamma$ of ScIrP are estimated to be 0.372(3) mJ K$^{-4}$ mol$^{-1}$ and 4.11(2) mJ K$^{-2}$ mol$^{-1}$, respectively, by the linear fit of the 2–3.3 K data.[18] Thus, the experimentally obtained $\gamma$ of ScRhP is almost twice as large as that of ScIrP.

This trend of the $\gamma$ values is consistent with the magnetization data. Figure 4(b) shows the temperature dependence of the magnetic susceptibility $\chi$ of ScRhP and ScIrP polycrystalline samples measured at a magnetic field of 1 T. Both the ScRhP and ScIrP data exhibit a weak temperature dependence except for a Curie tail, suggesting that the Pauli paramagnetic susceptibility is dominant in the $\chi$ data. Although the quantitative estimation of the Pauli paramagnetic contribution is difficult because of the difficulty in estimating the diamagnetic contribution of core electrons and the possible presence of a finite van Vleck susceptibility particularly for the Ir compound, the larger $\chi = 3.7 \times 10^{-5}$ cm$^3$ mol$^{-1}$ at 300 K for ScRhP than $\chi = 2.0 \times 10^{-5}$ cm$^3$ mol$^{-1}$ for ScIrP is consistent with the larger $\gamma$ of ScRhP

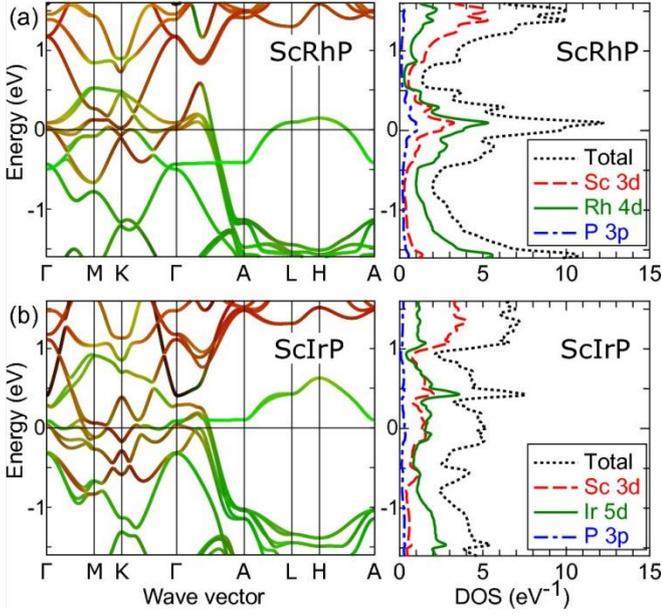

Fig. 6. Electronic states of (a) ScRhP and (b) ScIrP with spin–orbit interaction. The electronic band structure (left) and partial electronic density of states (right) are shown. The Fermi level is set to 0 eV.

than that of ScIrP.

The experimentally obtained $\gamma$ values of the ScRhP and ScIrP samples are almost the same as the calculated Sommerfeld coefficients $\gamma_{band}$. The electronic density of states at the Fermi level of $D(E_F) = 9.58$ and 5.16 states eV$^{-1}$ for ScRhP and ScIrP, respectively, estimated by first principles calculations with spin–orbit interaction, shown in the right panel of Fig. 6, yield $\gamma_{band} = 7.53$ and 4.05 mJ K$^{-2}$ mol$^{-1}$. Note that the effect of spin–orbit interaction on $D(E_F)$ of ScRhP and ScIrP is smaller than that of the typical noncentrosymmetric $5d$ electron systems LaPt$_3$Si, LaIrSi$_3$, and LaIrGe$_3$.[25–27] The $D(E_F)$ values of ScRhP and ScIrP are estimated to be 9.61 and 4.99 states eV$^{-1}$, respectively, using calculations without spin–orbit interaction. The comparisons between $\gamma$ and $\gamma_{band}$ in ScRhP and ScIrP indicate that the enhancement of $\gamma$ due to the electron–electron and electron–phonon interactions is small in both compounds. Hence, the almost double $\gamma$ of ScRhP compared with that of ScIrP strongly suggests that the $D(E_F)$ of the former is significantly larger than that of the latter.

We now examine the origin of the difference in $D(E_F)$ between ScRhP and ScIrP. As shown in the right panel of Fig. 6(a), the $E_F$ of ScRhP is located at a sharp peak of $D(E)$. This peak becomes broader in ScIrP, resulting in the smaller $D(E_F)$ in ScIrP than in ScRhP. The partial density of states for individual atoms, shown in the right panel of Fig. 6, indicates that the $D(E_F)$ mainly consists of the contributions of the Rh/Ir $4d/5d$ and Sc $3d$ orbitals. In ScRhP, both contributions are larger than those in ScIrP and the contribution of the Rh $4d$ orbitals is significantly enhanced compared with that of the Ir $5d$ orbitals in ScIrP.

As seen in the electronic energy dispersions of ScRhP and ScIrP shown in the left panel of Fig. 6, the energy bands in the $k_z = 0$ plane are narrower for ScRhP than for ScIrP. These differences probably reflect the fact that the $4d$ orbitals of a Rh atom are more localized than the $5d$ orbitals of an Ir atom, and that the orbital overlapping in the $ab$-plane in ScRhP is weaker than that in ScIrP owing to the larger $a$. As a result, the nearly flat bands around the L and H points come close to $E_F$ in ScRhP, giving rise to the significantly larger $D(E_F)$ in ScRhP than in ScIrP.

However, the $T_c = 2$ K of the ScRhP sample is significantly lower than $T_c = 3.4$ K of the ScIrP sample. The Debye temperatures of ScRhP and ScIrP are estimated to be $\theta_D = 287$ and 250 K, respectively, from the coefficient of the lattice heat capacity $\beta$, giving rise to a slightly higher phonon frequency $\omega_p$ in ScRhP than in ScIrP. This result is consistent with the smaller weight of the Rh atom than that of the Ir atom. Moreover, the electron–phonon interactions in both compounds are suggested to be weak from the fact that the experimentally obtained $\gamma$ values are almost the same as $\gamma_{band}$ estimated from the first principles calculations. If the conventional phonon-mediated superconductivity is realized in both compounds, the larger $D(E_F)$, the higher $\omega_p$, and the stronger electron–phonon interaction will give rise to a higher $T_c$. The reason why the $T_c$ in ScRhP is suppressed to be almost half of that in ScIrP despite the presence of the larger $D(E_F)$ and the higher $\omega_p$ is still open to future investigations. There is a possibility that some kind of pair breaking occurs in ScRhP or the $T_c$ in ScIrP is enhanced for some reason other than the usual electron–phonon interactions.

## 4. Conclusions

We synthesized polycrystalline samples of ScRhP, which is a new member of the ordered Fe$_2$P-type pnictide family. ScRhP is found to show a bulk superconducting transition at $T_c = 2$ K, through electrical resistivity, magnetization, and heat capacity measurements. The upper critical field at $T = 0$ is estimated to be $\mu_0 H_{c2}(0) = 0.5$ T from the magnetic field dependence of $T_c$. The $T_c$ of ScRhP is almost half of that of ScIrP, while the $H_{c2}(0)$ of the former is almost 1/10 of that of the latter. Since ScRhP has a larger $D(E_F)$ and a higher $\omega_p$ than ScIrP, the lower $T_c$ in ScRhP may not be simply explained in cases where the conventional phonon-mediated superconductivity is realized in both compounds. However, the much higher $H_{c2}(0)$ of ScIrP than that of ScRhP implies that the $H_{c2}(0)$ of ScIrP is significantly enhanced by the strong antisymmetric spin–orbit interaction of the Ir $5d$ electrons in the noncentrosymmetric crystal structure.

**Acknowledgments**

This work was partly carried out at the Materials Design and Characterization Laboratory under the Visiting Research Program of the Institute for Solid State Physics, University of Tokyo and partly supported by JSPS KAKENHI Grant Numbers 25800188 and 16H01072. We are grateful to D. Hirai and K. Nawa for their support in the low-temperature experiments using a $^3$He refrigerator, to T. Yamauchi for his help in the magnetization measurements, and to A. Yamakage for helpful discussions.

________________________________________________

*yokamoto@nuap.nagoya-u.ac.jp